\documentclass[twocolumn]{aastex631}
\usepackage{subfig}
\hypersetup{
    colorlinks=true,
    linkcolor=blue,
    filecolor=magenta,      
    urlcolor=blue,
    pdftitle={Overleaf Example},
    pdfpagemode=FullScreen,
    }

\newcommand{\sqcm}{cm$^{-2}$\,}  
\newcommand{\kms}{$\rm km\, s^{-1}$\,} 
\newcommand{\lya}{Ly$\alpha$\,}
\newcommand{\lyb}{Ly$\beta$}

\newcommand{\HI}{\mbox{H\,{\sc i}}}

\newcommand{\CIV}{\mbox{C\,{\sc iv}}}

\newcommand{\Rvir}{$R_{\rm vir}$}
\newcommand{\ew}{$\rm EW_0$}
\newcommand{\logN}{$\log_{10} N/{\rm cm^{-2}}$} 
\newcommand{\Msun}{M$_{\odot}$}
\newcommand{\ang}{\mbox{\normalfont\AA}}
\newcommand{\NHI}{$N_{\rm H \textsc{i}}$}
\newcommand{\fc}{$f_{\rm c}$}
\newcommand{\Rev}[1]{\textcolor{black}{#1}}

\shorttitle{Neutral gas surrounding $z\approx3$ LAEs}
\shortauthors{Banerjee et al.}

\begin{document}

\title{MUSEQuBES: Connecting \HI\ absorption with Ly$\alpha$ emitters at $z \approx 3.3$}

\correspondingauthor{Eshita Banerjee, Sowgat Muzahid}
\email{eshitaban18@iucaa.in, sowgat@iucaa.in}

\author[0009-0002-7382-3078]{Eshita Banerjee}
\affiliation{IUCAA, Post Bag 04, Ganeshkhind, Pune, India, 411007} 

\author[0000-0003-3938-8762]{Sowgat Muzahid}
\affiliation{IUCAA, Post Bag 04, Ganeshkhind, Pune, India, 411007} 


\author{Joop Schaye}
\affiliation{Leiden Observatory, Leiden University, P.O. Box 9513, NL-2300 AA Leiden, the Netherlands}

\author{Jérémy Blaizot}
\affiliation{Centre de Recherche Astrophysique de Lyon UMR5574, Univ Lyon, CNRS, F-69230 Saint-Genis-Laval, France}

\author{Nicolas Bouché}
\affiliation{Universit\'e Lyon 1, CNRS, Centre de Recherche Astrophysique de Lyon (CRAL), Saint-Genis-Laval, France}

\author{Sebastiano Cantalupo}
\affiliation{Department of Physics, University of Milan Bicocca, Piazza della Scienza 3, I-20126 Milano, Italy}

\author{Sean D. Johnson}
\affiliation{Department of Astronomy, University of Michigan, 1085 S. University, Ann Arbor, MI 48109, USA}

\author{Jorryt Matthee}
\affiliation{Institute of Science and Technology Austria (ISTA), Am Campus 1, Klosterneuburg, Austria}

\author{Anne Verhamme}
\affiliation{Observatoire de Gen\`eve, Universit\'ee de Gen\`eve, 51 Ch. des Maillettes, CH-1290 Versoix, Switzerland}





\begin{abstract}

\noindent
We present a comprehensive analysis of \HI\ absorption around 96 \lya\ emitters (LAEs) at $z\approx3.3$ (median \lya\ luminosity $\approx10^{42}$ erg.s$^{-1}$). These LAEs were identified within 8 MUSE fields, each $1'\times1'$ on the sky and centered on a bright background quasar, as part of the MUSEQuBES survey. Using Voigt profile fitting for all \HI\ absorbers detected within $\pm500$~\kms\ of these LAEs, we compiled a catalog of 800 \HI\ absorption components. Our analysis shows that \HI\ absorption is enhanced near the LAEs compared to the IGM. However, no trend is found between the column densities of \HI\ absorbers and their impact parameters from the LAEs (spanning $\approx54$–260 pkpc). Additionally, all galaxies associated with Lyman-limit systems have impact parameters $>50$~pkpc from the quasar sightlines, suggesting that true absorber-hosts may be too faint to detect. The LAEs show an overall \HI\ covering fraction (\fc(\HI)) of $\approx88\%$ for a  threshold \logN(\HI)$~=15$. Notably, at the same threshold, the pairs/group LAEs exhibit a $100\%$ \HI\ covering fraction out to $\approx 250$~pkpc. In contrast, isolated LAEs consistently show a lower \fc(\HI) of $\approx80\%$. This environmental influence on \fc(\HI) is also evident up to $\approx 300$~\kms\ in differential bins of line-of-sight velocity. We find an anti-correlation between \fc(\HI) and the rest-frame \lya-emission equivalent width (\ew). Based on the \lya-shell model, this could imply that gas-rich galaxies tend to reside in gas-rich environments or that the higher \ew\ LAEs are more efficient at ionizing their surrounding medium. 

\end{abstract}



\section{INTRODUCTION}

The circumgalactic medium (CGM) is a dynamic, complex, multiphase gaseous region, extending out to $\sim$few~100~kpc, enveloping galaxies \citep[see the review by][]{Tumlinson2017}. It contains the imprints of crucial processes such as galactic-scale winds and the accretion of intergalactic baryons via the ``hot mode''/``cold mode'' along cosmic filaments \citep{Keres_2005, van_de_voort_2011a}. This cyclic interplay of baryons in the CGM, often referred to as the ``baryon cycle'', is a key factor governing galaxy evolution \citep[e.g.,][]{Peroux2020}. It is widely believed that these processes played a crucial role in determining the evolution of the cosmic star formation rate density (SFRD) which shows a peak at redshift of $z \approx 2-3$ and declines by almost an order of magnitude at both higher and lower redshifts \citep[e.g.,][]{Madau2014review, van_de_voort_2011b}. Given that hydrogen accounts for the majority of the baryonic matter in the universe, it has become a vital tool for probing the cool gas that acts as the fuel for star formation. 


Probing the cool gas in the CGM in emission is challenging owing to its low density ($n_{\rm H} \sim 10^{-3}\, \rm cm^{-3}$ at $z\approx 3$). State-of-the-art integral field spectrographs (IFS) such as VLT/MUSE \cite[Multi-Unit Spectroscopic Explorer;][]{bacon2010muse} and Keck/KCWI \cite[Keck Cosmic Web Imager;][]{Morrissey_2018} have enabled the exploration of the diffuse medium surrounding galaxies in \lya\ emission within a few tens of kpc from normal galaxies \citep[e.g.,][]{Wisotzki2018, Erb2018, Guo2024a, Guo2024b} and a few hundreds of pkpc around active galaxies \citep[e.g.,][]{Cantalupo2014,Borisova_2016, Fabrizio2019}. Extended \lya\ and metal line emission is also detected in several individual cases and in stacks \citep[e.g.,][]{Steidel2011, Zabl_2021,Johnson_2022,Leclercq2022,Rdutta2023,Guo2023_stack,Liu2024, Johnson2024}. However, the absorption spectra from bright background sources, such as quasars, remain the most sensitive tool to probe this elusive medium.

Historically, the relation between galaxies and the neutral gas surrounding them was studied by identifying host-galaxies of strong \HI\ absorbers, such as Lyman-limit systems \citep[LLSs; \NHI$>10^{17.2}$~\sqcm; e.g.,][]{Lofthouse_2023}, and  Damped Lyman-$\alpha$ absorbers \citep[DLAs; \NHI$\gtrsim10^{20.3}$~\sqcm; e.g.,][]{Peroux_2012, Krogager_2017, Mackenzie_2019}, seen in quasar spectra. Several investigations have primarily focused on high-metallicity DLAs, driven by the presumption that these DLAs possess more luminous galaxy counterparts, owing to the underlying mass-metallicity relation \citep[e.g.,][] {Neeleman_2013}. These studies utilized facilities such as VLT/X-Shooter \citep[e.g.,][]{Krogager_2017}, VLT/SINFONI \citep[e.g.,][]{Peroux_2012}, and ALMA \citep[e.g.,][]{Neeleman2017, Neeleman2019}, revealing close associations (often within a few tens of kpc) between DLAs and their host galaxies. However, recent studies with state-of-the-art IFSs such as the MUSE with a field of view (FoV) of $1'\times1'$, revealed that galaxies associated with the DLAs/LLSs are often located at larger impact parameters \citep[$>50$~pkpc; see,][]{Mackenzie_2019, Lofthouse_2023}. This phenomenon could potentially be attributed to the fact that the true host of the absorber remains undetected owing to its relative faintness \citep[e.g.,][]{Rahmati_schaye2014}.

Recently, using MUSE observations, \citet{Lofthouse_2023} reported a detection rate of $\approx80$ percent for \lya\ emitting galaxies (LAEs) around absorbers with \NHI~$>10^{16.5}$~\sqcm\ with their impact parameter ranging from 18--275 pkpc. Based on the higher number density of these LAEs compared to field galaxies, they concluded that LAEs are significantly clustered around high \HI\ column density absorbers. Besides, they found that LAEs in groups exhibit a three times higher covering factor of optically thick gas compared to isolated systems. The absence of a substantial correlation between emission and absorption characteristics led them to conclude that the CGM is patchy and inhomogeneous. They argued that very large samples of quasar-galaxy pairs are required to reveal the signatures of the baryon cycle. While high-column density \HI\ absorbers serve as reliable indicators of galaxy overdensities, it is crucial to investigate the connection(s) between galaxies and gas in an ``absorption-blind'' manner.

\citet{Rudie_2012}, as a part of the Keck Baryonic Structure Survey \citep[KBSS; see also, ][]{Adelberger_2005, Steidel_2010, Rakic_2012, Rudie_2012, Rudie_2013, Rudie_2019, Turner_2014, Chen2020}, performed a complete Voigt profile decomposition of all the \HI\ absorbers down to \NHI~$\sim10^{12}$~\sqcm\ around a sample of Lyman-break galaxies (LBGs) and found a strong enhancement of the \HI\ absorption relative to randomly located regions. This enhancement extends out to $\approx 300$~pkpc in the transverse direction and $\pm 700$~\kms\ along the line of sight (LOS). \Rev{The halo masses of the LBGs in their sample are $\sim 10^{12}$~\Msun.} Clearly, these are massive structures and represent only the ``tip of the iceberg'' of the high-$z$ galaxy population. It is important to investigate the connections between `typical' high-$z$ galaxies and neutral gas around them. The \lya\ emitting galaxies with halo masses of $\sim 10^{11}$~\Msun\ are a good candidate for low-mass, high-$z$ galaxies \citep[]{Herrero_Alonso_2021}. Such low-mass galaxies are more susceptible to galactic-scale outflows, owing to their shallow gravitational potentials. Mapping the distribution of gas and metals around LAEs is therefore crucial to exploring the galaxy-CGM connections in the unexamined low-mass regime.

The MUSE Quasar-field Blind Emitters Survey \citep[MUSEQuBES;][]{Muzahid_2020, Muzahid_2021, Banerjee_2023} is dedicated to investigating the correlation between gas properties and low-mass galaxies at redshifts $z\approx 3-4$. The MUSEQuBES galaxies, identified as LAEs, have a median stellar mass $M_* \approx 10^{8.9}$~\Msun. The MUSEQuBES galaxy sample is obtained without any prior knowledge of the presence or absence of gas (it is `absorption-blind'). \citet{Muzahid_2021} carried out a spectral stacking analysis of the CGM absorption around $96\, z\approx3.3$ LAEs and found that the median rest-frame equivalent width ($W$) of \HI\ absorption correlates with the SFR and environment of the LAEs. For the same LAE sample, \citet{Banerjee_2023} studied the connection between galaxies and \CIV\ absorbers at $z\approx3.3$. In continuation of the MUSEQuBES survey, here we present a thorough Voigt profile decomposition of \HI\ absorbers detected within $\pm500$~\kms\ LOS velocity windows around each of the 96 MUSEQuBES LAEs. 

There is no consensus on the extent of the CGM from its host galaxy. Some low-$z$ surveys suggest that it extends out to the \Rvir\ of galaxies \citep[e.g.,][]{Tumlinson2017, sayak2024}. In our current work, we are probing a large distance of up to $\approx 320$~pkpc from these galaxies, much larger than their typical \Rvir\ ($\approx42$~pkpc). Therefore, it is more reasonable to say that we are probing the extended gaseous medium associated with these LAEs. However, for simplicity, here we will call it the CGM.


This paper is organized as follows: In Section~\ref{sec:data}, we provide a concise overview of the quasar and galaxy data employed in this paper, with particular emphasis on the preparation of the \HI\ absorption line catalog. In Section~\ref{sec:results}, we present our analysis and results. In Section~\ref{sec:discussions}, we discuss and interpret our results. Finally, a brief summary is presented in  Section~\ref{sec:summary}.

Throughout this study, we adopt a flat $\Lambda$CDM cosmology with $H_0$ = $70\,\rm km\, s^{-1}\, Mpc^{-1}$, $\Omega_{\rm M} = 0.3$ and $\Omega_\Lambda = 0.7$, and all distances given are in physical kpc (hereafter, pkpc) unless specified otherwise.

\section{Data and measurements} 
\label{sec:data} 

The quasar and galaxy data presented in this study is drawn from the MUSEQuBES survey. Our survey contains excellent quality optical spectra of 8 UV-bright background quasars, obtained using the VLT/UVES and Keck/HIRES spectrographs with a spectral resolution of $R \approx 45,000$. All spectra exhibit a signal-to-noise ratio (SNR)~$\gtrsim30$ per pixel within the \lya\ forest, except for quasar BRI1108-07, which has an SNR of $\approx15$. Further details on the data and quasar continuum fitting can be found in section~2 of \citet{Muzahid_2021}. The subsequent subsections contain brief discussions on the galaxy data and details of absorption line measurements.

\begin{table*}
\begin{center} 
\caption{The best-fitting parameters obtained from  Voigt profile fitting of the \HI\ absorbers.}
\begin{tabular}{lccccccc}
    \hline
    QSO &        $z$ &          $z$-error &         $b$~(\kms) &      $b$-error~(\kms) &  \logN &      \logN-error & Detection ID \\
    \hline
Q1422+23  &   2.958128 & 2.8e-06  & 23.00 &  0.35  &  13.52 & 0.01 & 0\\
...       &    ...     &    ...   &  ...  &  ...   &  ... & ... & ... \\ 
BRI1108-07 & 3.584494 &  8.8e-06 & 28.95 &  3.46 &    17.25 &  0.80 &   1   \\ 
...       &    ...     &    ...   &  ...  &  ...   &  ... & ... & ... \\ 
    \hline
    \end{tabular}
    \label{tab:catalog}
    \vskip0.2cm 
    \end{center}
Notes-- Detection $\rm ID = 1$: lower limit on \NHI; $\rm ID=0$: well-constrained \NHI. The complete version of this table is available online. \\
\end{table*}

\subsection{LAE Sample} 

A total of 50h guaranteed time MUSE observations (PI: J. Schaye) of 8 quasar fields allowed us to identify 96 LAEs at $2.9<z<3.8$, spanning an impact parameter ($\rho$) range of $16<\rho<320$~pkpc. These LAEs were detected solely based on their \lya\ emission lines, which typically exhibit offsets of hundreds of \kms\ from the galaxies' systemic redshifts \citep[see][]{Steidel_2010, Rakic+11, Shibuya_2014, Verhamme18}. An empirical relation between this offset and the Full Width at Half Maximum (FWHM) of the \lya\ emission peak was provided for our sample by \cite{Muzahid_2020} i.e., $V_{\rm offset} = 0.89\times {\rm FWHM} -58$~\kms. The systemic redshifts of the LAEs in this study are derived using this relation.

The median stellar mass for our sample is $M_*$~$=10^{8.9}$~\Msun\ and median SFR is $\approx 1.2$~\Msun$/\rm year$. Note that these SFR values are not corrected for dust and are derived from the measured UV luminosity density values using the local calibration relation of \cite{Kennicutt_1998} corrected to the \cite{Chabrier_2003} initial mass function. The stellar masses of these LAEs are estimated from the measured SFR values, assuming the star-forming main sequence relation of \citet{Behroozi2019}. Although there are several assumptions involved in estimating these values, they are very close to the stellar mass and SFR of typical LAE populations at these redshifts, i.e., $M_* = 10^8 - 10^{8.5}$~\Msun\ and an SFR of $\approx 2.3$\Msun$/\rm year$, as estimated based on broader SED fitting and rest-optical emission line studies \citep[see,][]{Ono2010, Trainor2015}. Further, using the halo abundance matching relation from \citet{moster2013galactic}, we estimated the halo mass ($M_{\rm vir}$) and corresponding virial radius (\Rvir) {for a subset of 39 out of 96 LAEs, where the SFR could be measured from their UV continuum. For the remaining LAEs, the UV continuum was either undetected with $>5\sigma$ confidence or blended with foreground sources \citep[see,][]{Muzahid_2020}.} The halo masses (virial radii) range from $10^{11.0}-10^{11.7}$~\Msun\ ($33-59$~pkpc) with a median value of $10^{11.3}$~\Msun\ ($42$~pkpc). {The $5^{\rm th}$ to $95^{\rm th}$ percentiles of the \lya\ luminosity for the MUSEQuBES LAEs are $\log_{10} (L_{\rm Ly\alpha}/ {\rm erg\, s^{-1}}) = 41.5$ and $42.7$; the corresponding \lya\ line fluxes  are $3\times10^{-18}$ and $ 4\times10^{-17}$~{$\rm erg~cm^{-2}~s^{-1}$}.} Here we focus on the \HI\ absorbers located within $\pm500$~\kms\ of the corrected LAE redshifts. Further details regarding the MUSE data analysis and galaxy properties can be found in \cite{Muzahid_2021}.



\subsection{\HI\ absorption line measurements}    

The accurate measurement of \HI\ column densities in the spectra of high-$z$ quasars is challenging due to (a) the blending of \HI\ absorption with other \HI\ or metal lines present in the \lya\ forest and (b) the saturation of strong \HI\ absorption components. To overcome these challenges, it is imperative to simultaneously fit all the higher-order Lyman-series lines. This approach not only facilitates the accurate measurement of column density but also aids in effectively identifying contamination, thereby simplifying its removal.

In our study, we conducted a detailed Voigt profile decomposition of \HI\ absorption detected within $\pm500$~\kms\ of the 96 MUSEQuBES LAEs. Note that a given absorber may be associated with multiple LAEs. For Voigt profile decomposition, we used the {\sc vpfit} code \citep{Carswell2014} that works based on $\chi^2$-minimization. {\sc vpfit} allowed us to simultaneously fit the available Lyman-series lines as well as the contaminating metal lines. We obtained the best-fitting values for the redshift, Doppler parameter ($b$), and column density ($N$) of the $800$ associated individual \HI\ components in the redshift range $2.914 \leq z \leq 3.823$ (median $z=3.364$). The best-fitting parameters are summarized in Table~\ref{tab:catalog}. To ensure the reliability of our measurements, we manually examined the fitted profiles and repeated the fitting process until a satisfactory fit was achieved, in addition to the $\chi ^2$ and AICc (Akaike information criterion; corrected for small sample sizes) values of the model.


\begin{figure}
    \centering
    \includegraphics[width=\linewidth]{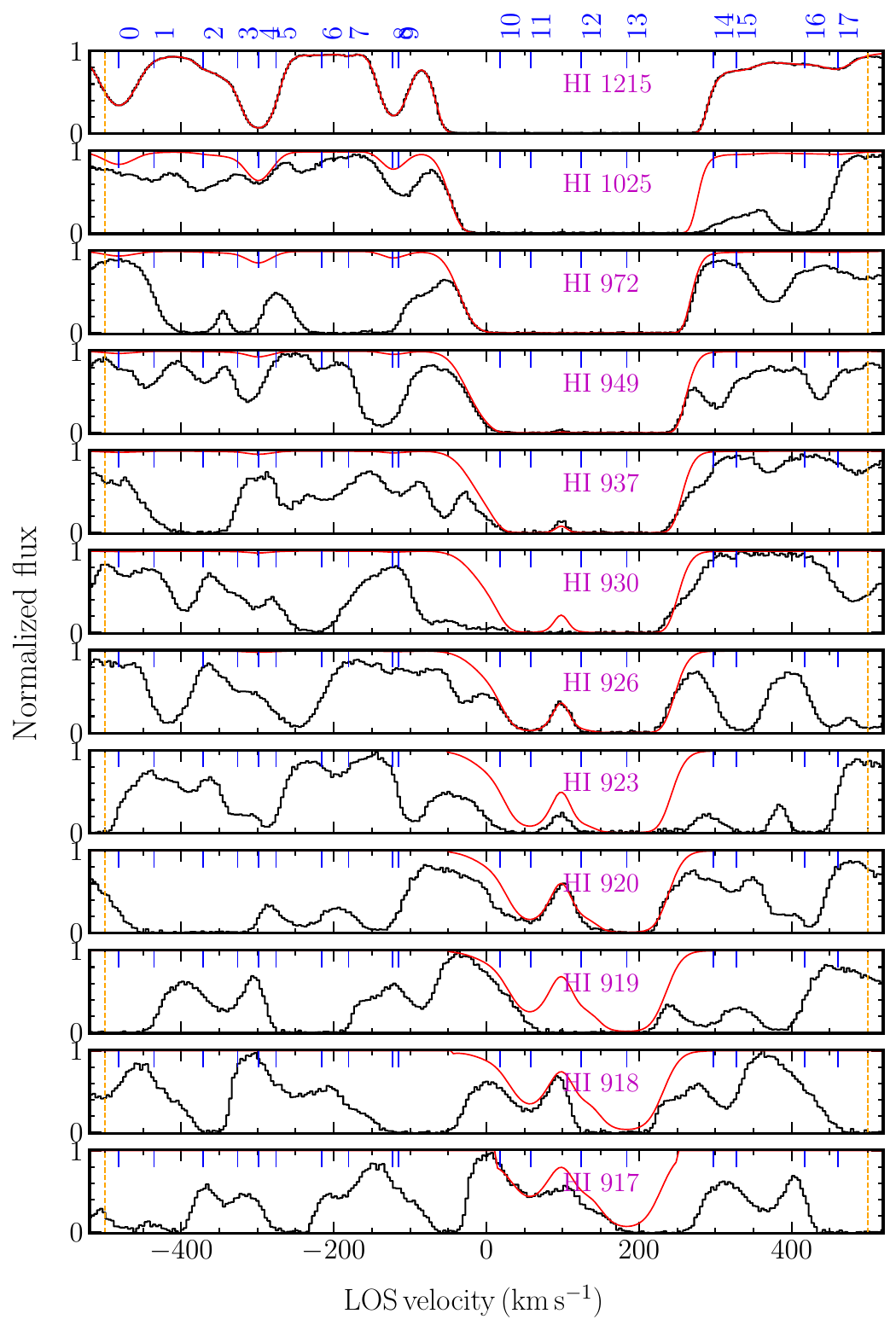}
    \caption{\HI\ absorption profiles associated with one of the LAEs towards Q1422+23. The normalized quasar spectra are represented in black, with the best-fitting Voigt profiles shown in red. The LOS velocity of $0$~\kms\ denotes the systemic redshift of the corresponding LAE. The $\pm500$~\kms\ velocity range is depicted by the orange-dashed vertical lines. The centroids of individual Voigt profile components within this velocity window are indicated by blue ticks. The Lyman-series lines are plotted from top to bottom in descending order of wavelength, as mentioned in each panel. The column density is well constrained for the unsaturated \lya\ components (i.e., 0--3, 5--9, and 14--17) and for the components with unsaturated high-order lines (i.e., 10--12). All of these components are classified as $\rm ID=0$ absorbers in Table~\ref{tab:catalog}. On the other hand, all the higher-order lines corresponding to component 13 are saturated. We therefore treat the {\sc vpfit} returned column density for this component as a lower limit (i.e. $\rm ID=1$). The total $N(\HI)$ within $\pm500$~\kms\ of this LAE is \logN~$=17.1$, which is also treated as a lower limit. The unfitted absorptions are contaminating lines originating from other redshifts.
    }   
\label{fig:vpfit_exmp}
\end{figure}

\begin{figure}
\includegraphics[width=0.48\textwidth]{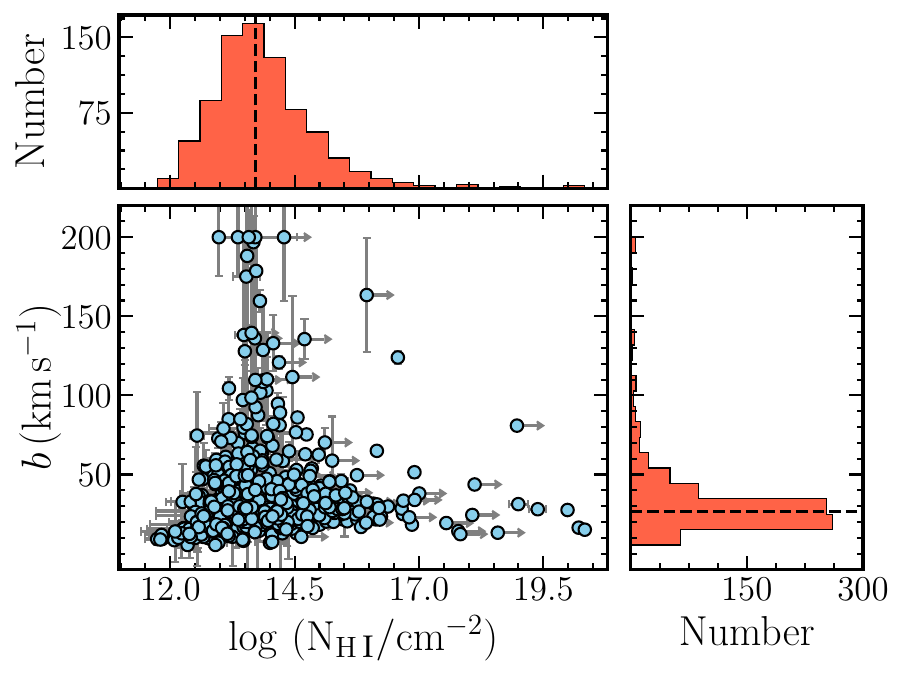}
\caption{Scatter plot of $b$ vs. $N$ for all the 800 \HI\ absorption components detected within $\pm500$~\kms of the LAE systemic redshifts. The points marked with an rightward arrow indicate the components for which $N(\HI)$ cannot be measured due to blending and/or saturation. The side panels present the distributions of the $N$ and $b$ values, with the black dashed lines indicating their respective median values.}  
\label{fig:bn_scatter}
\end{figure}

Due to blending, the presence of higher-order Lyman-series lines does not always guarantee a well-resolved structure for heavily saturated absorption profiles. In such instances, we had to rely mostly on the saturated \lya\ line. Previous studies \citep[e.g.,][]{Shull2000, Songaila2001, Danforth_2010} have indicated that, when using a single-line measurement of \HI, the fitted profile tends to be $\approx 1.5$ times broader than what is estimated through curve-of-growth (COG) measurements derived from higher-order Lyman-series lines. An overestimate of the Doppler width leads to an underestimate of the column density. As a result, the column density estimate for these components represents a lower limit. Within a velocity range of $\pm500$~\kms\ from the LAEs, we have identified 66 components (comprising $8\%$ of the total number of components) that are treated as lower limits in our analysis. These components either suffer from saturation without the availability of other Lyman-series lines or, if available, the true velocity component structure cannot be resolved due to blending. They are classified as detection $\rm ID=1$ absorbers in Table~\ref{tab:catalog}.

Our absorber catalog contains 48 weak, un-saturated \lya\ absorption components, comprising $6\%$ of the total components, for which no higher-order lines are available. Since the lines are deemed to be unsaturated, obtaining accurate column densities is not problematic for these components \citep[see also][]{Kim2021}. Hence, these absorbers, along with the rest of the absorbers for which we can constrain the \NHI\ value using the unsaturated higher-order Lyman series lines, are classified as $\rm ID=0$ components.

Fig.~\ref{fig:vpfit_exmp} shows an example of our Voigt profile decomposition. Each panel displays the spectra (in black) in LOS velocity space along with the best-fitting profiles (in red) for an LAE towards the Q1422+23 sightline. The $0$~\kms\ velocity corresponds to the systemic redshift of the LAE. The centroid of each \HI\ component within $\pm500$~\kms\ of the LAE (region marked by orange dashed vertical lines) is indicated by blue ticks. All components, except number 13, are classified as $\rm ID=0$. Owing to the saturation of all the identified higher-order lines, only a lower limit on \NHI\ can be derived for the component labeled 13, which is classified as an $\rm ID=1$ absorber.

Fig.~\ref{fig:bn_scatter} shows a scatter plot of $b$(\HI) vs.\ $N$(\HI) of all 800 \HI\ components in our catalog with arrows indicating lower limits on $N$(\HI). The distributions of these two parameters are shown in the side panels. The absorber sample spans the range \logN\ $=11.7-20.3$ with a mean (median) of $=13.9$ ($13.7$). For components with \logN\ $<16$, the median error in column density is 0.05 dex, while for components with \logN\ $>16$, it is 0.04 dex. The mean (median) value of the Doppler parameter is 34~\kms\ (27~\kms). Only a small fraction of the components show very narrow ($b\approx 6.6$~\kms; $\approx0.6$\%) or very broad ($b>100$~\kms; $\approx3$\%) profiles. We verified that, except for a few cases, the large $b$-value components arise in blends and are susceptible  to continuum fitting uncertainties. We find that some of the narrow components can be attributed to unidentified low-$z$ metal lines. hence, they were excluded from our catalogue.


\section{Analysis and Results}
\label{sec:results}

We will now proceed with the analysis of the overall properties of the \HI\ absorbers, which are located within $\pm500$~\kms\ of the MUSEQuBES LAEs. This analysis will involve comparing the characteristics of these absorbers to the typical absorption features observed in the IGM at a similar redshift.

\subsection{\HI\ column density vs. impact parameter} 
\label{sec:column_profile}
\begin{figure*}
   \includegraphics[width=\textwidth] {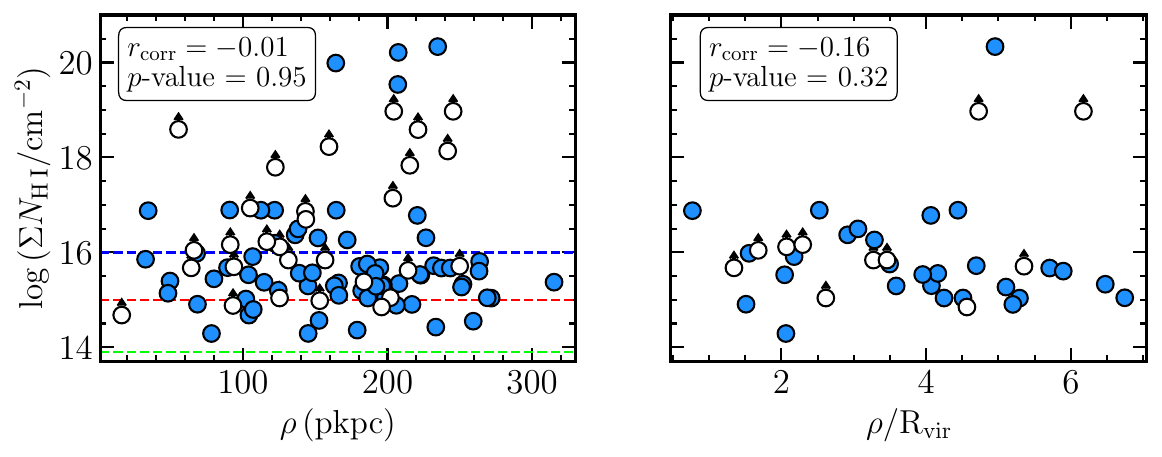}
    \caption{ {\tt Left:} The total \NHI\ ($\Sigma$\NHI) within $\pm500$~\kms\ of the LAEs against the impact parameter ($\rho$). The data points represented by empty circles with upward arrows denote lower limits on \NHI, indicating that these data points have some contribution from unresolved/blended \HI\ component(s). The average value of the total \NHI\ within $\pm500$~\kms\ of the LAEs is \logN$>16.0$ (indicated by the blue dashed line), where the inequality is due to the presence of unresolved lower limits in our data. The Spearman correlation results, mentioned in the plot, reveal the absence of any significant trend between $\Sigma$\NHI\ and $\rho$. \Rev{The green and red dashed line in the plot represents the \NHI\ corresponding to overdensities of $\delta \approx 1$ and $\approx10$ respectively,} suggesting that in the vicinity of the LAEs, there is an enhancement of neutral hydrogen. \Rev{{\tt Right:} Similar to the left panel, but here the impact parameter is normalized by $R_{200}$ or \Rvir. Only LAEs with a continuum detected at $>5\sigma$ are shown.} }
    \label{fig:col_profile}
\end{figure*}

First, we focus on the distribution of \HI-absorbing gas surrounding the LAEs. Fig.~\ref{fig:col_profile} (left panel) shows the distribution of the total column density ($\Sigma$\NHI) calculated within a velocity range of $\pm500$~\kms\ from the LAEs as a function of the impact parameter ($\rho$) of the LAEs. The empty circles represent lower limits on \NHI\ for the LAEs that possess at least one $\rm ID=1$ component within $\pm500$~\kms. The plot reveals a lack of any noticeable correlation between the total \HI\ column density and $\rho$. This observation is further confirmed by a Spearman correlation test with $\rm r_{corr}=-0.01$ and a $p$-value of $0.95$. When we exclude the lower limits, the results of the Spearman correlation test yield $\rm r_{corr}=-0.07$ and $p$-value~$=0.54$ indicating no significant correlation. \Rev{In right panel, we are showing the same plot, but here the impact parameter is normalized by the \Rvir, considering only those LAEs with a continuum detected at $>5\sigma$ confidence. Here, also, we do not see any significant correlation.}

\Rev{ The mean value of total \NHI\ within $\pm500$~\kms\ of the LAEs is $\approx10^{16.0}$~\sqcm\ (blue dashed line in the left panel of Fig.~\ref{fig:col_profile}), which may be an underestimate due to the presence of several lower limits on \NHI. This is quite enhanced compared to the IGM, where the typical overdensity is $\approx 1$. In the same plot, the green (red) dashed line represents the \NHI~$=10^{13.9}\, (10^{15.0})$~\sqcm\ corresponding to overdensities of $\approx 1 \, (10)$ at $z\approx 3$ \citep[see][]{Schaye_2001}. Our observed values are generally higher, indicating that \HI\ absorption is elevated near the LAEs compared to random regions in the IGM at similar redshifts. We obtained consistent results when we used a smaller velocity window of $\pm250$~\kms\ to associate LAEs and \HI\ absorption.}


Finally, another intriguing aspect of Fig.~\ref{fig:col_profile} is that the majority of the absorbers with \logN $>17.2$ are associated with LAEs with large impact parameters, in contrast to conventional wisdom. Considering that excessive noise from quasars can result in the non-detection of LAEs at smaller impact parameters, note that a total of 96 LAEs are detected within the $320\, {\rm pkpc}\times320\, {\rm pkpc}$ FoV. Assuming a random distribution of galaxies, we expect approximately 2 LAEs within $42\, {\rm pkpc}\times42\, {\rm pkpc}$, where $42$~pkpc is the typical virial radius of the MUSEQuBES LAEs. In fact, we detect 3 LAEs within $42$~pkpc of the quasar sightline. This suggests that the non-detection of LAEs at smaller impact parameters is not because of some issue with the quasar PSF subtraction. We discuss these systems further in Section~\ref{sec:discussions}.

\subsection{\HI\ covering fraction around LAEs}

The CGM is known to be an inhomogeneous and patchy medium,  exhibiting variations in its physical and chemical properties across different regions. The covering fraction of atomic and ionic species within the CGM serves as a quantitative measure of this patchiness, providing valuable insight into the underlying processes that govern galaxy evolution. In this section, we focus on the \HI\ covering fraction, \fc(\HI), near the LAEs. For a given threshold column density, $N_{\rm Th}$, it is defined as\footnote{In this study, we computed \fc\ within a defined transverse distance i.e., \fc~$(<\rho)$, unless we are doing differential binning of impact parameter.}: 
\begin{equation}
f_{\rm c}(\HI) = \frac{n_{\rm Hit} (N \geq N_{\rm Th})}{n_{\rm Total}}~, 
\label{eq:fc}
\end{equation}
where $n_{\rm Hit}$ is the number of LAEs hosting a total \HI\ column density \NHI~$\geq N_{\rm Th}$ in their CGM, measured within a specific LOS velocity window (here, $\pm500$~\kms) around the LAE redshifts, and $ n_{\rm Total}$ is the total number of LAEs for which the quasar spectra have sufficient SNR to detect absorption down to the threshold column density of $N_{\rm Th}$.

\begin{figure}
\includegraphics[width=\linewidth]{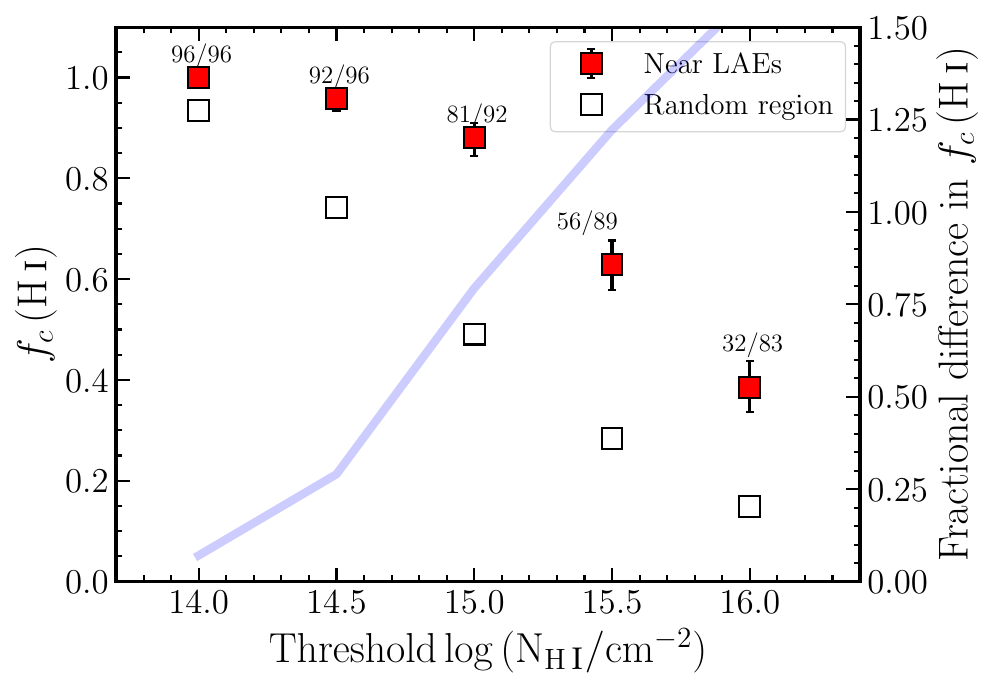}
\caption{The average \HI\ covering fraction for $\rho < 320$~pkpc   for different threshold $N(\HI)$ values. The red squares  represent the \fc\ estimated using \HI\ absorbers within $\pm500$~\kms\ of the LAEs, while the empty squares depict the \fc\ for the IGM (see text). The error bars on the red squares  indicate the $68\%$ Wilson-score confidence interval. The numbers adjacent to the squares correspond to the respective values of $n_{\rm Hit}$ and $ n_{\rm tot}$ as in Eqn.~\ref{eq:fc}. For all threshold column densities, \fc(\HI) is consistently significantly higher near LAEs compared to random locations. The line and the right y-axis represent the fractional difference in \fc(\HI) with respect to random regions, exhibiting a gradual increase with the threshold column density. 
}
\label{fig:fc}
\end{figure}

To determine the $3\sigma$ limiting column densities of our quasar spectra, we used the relation provided by \cite{Hellsten1998}, following a similar approach as described in section 3.1.1 of \cite{Banerjee_2023}. These limits are calculated using the median SNR in the \lya\ forest region for each spectrum, as reported in Table~2 of \citet{Muzahid_2021}. The undetected line is assumed to have a $b$ parameter of 35~\kms, which is the median value for the detected components. It is worth noting that the limiting column density is $\approx 10^{12.7}$~\sqcm\ for all 8 sightlines, which is lower than the threshold value we have used to calculate the \fc(\HI) in this paper. Additionally, we excluded the LAEs for which the lower limit on the detected \HI\ is lower than $N_{\rm Th}$, since it is not guaranteed that the total \NHI\ for such cases is higher than the threshold value.

\begin{figure*}
    \hbox{\includegraphics[width=0.50\textwidth]{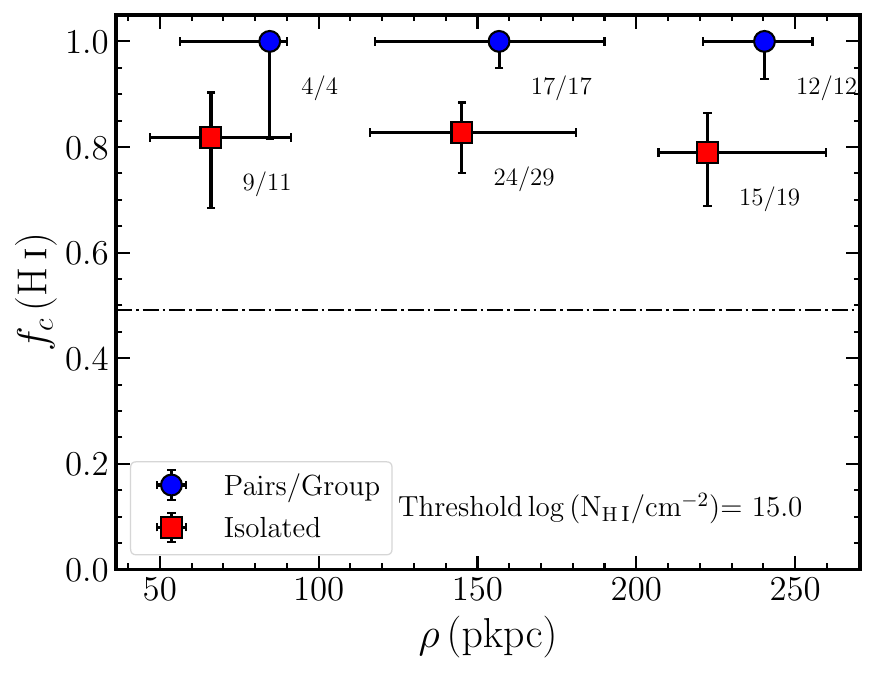} 
    \includegraphics[width=0.50\textwidth]{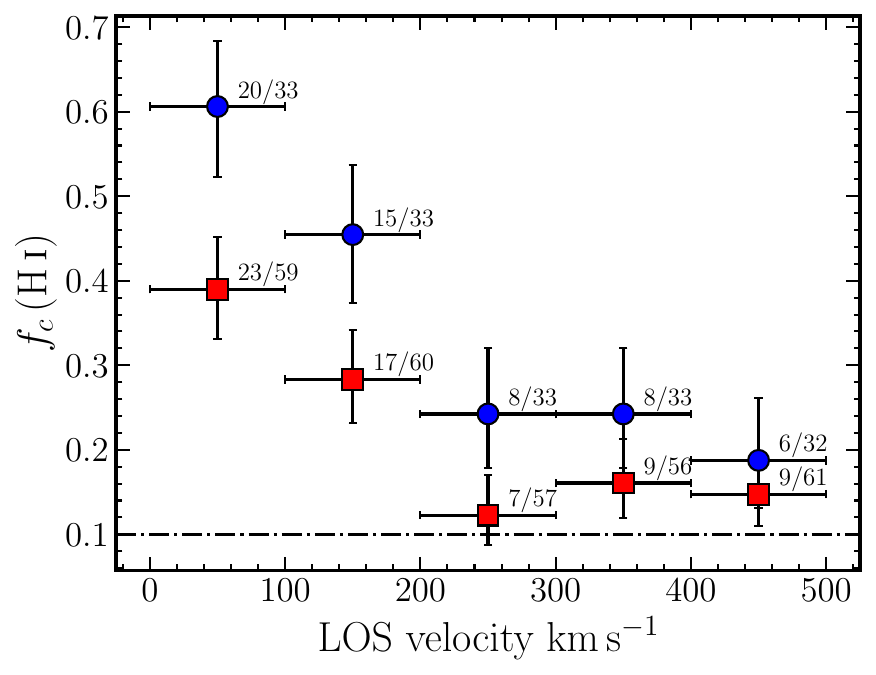}
    }
    \caption{{\tt Left:} The differential \HI\ covering fraction for a threshold \logN~$=15.0$ and velocity window $\pm500$~\kms\ for the isolated (red squares) and pairs/group (blue circles) LAEs against impact parameter. The error bars along the y-axis indicate the $68\%$ Wilson-score confidence intervals. Along the x-axis, the data points and error bars, respectively, represent the median values and the $16^{\rm th}-84^{\rm th}$ percentile range of the $\rho$ distribution within each bin. The black dash-dotted line shows \fc(\HI) for the IGM at the same threshold column density. Although no visible trend is observed with respect to $\rho$, the pairs/group LAEs consistently exhibit a higher \fc(\HI) across all three bins. {\tt Right:} \fc(\HI) with a differential LOS velocity window used to associate the \HI\ absorbers with the LAEs. The x-axis denotes the midpoint of each velocity bin, with error bars representing the span of that bin. As in the {\tt left} panel, the threshold column density used here is \logN$=15.0$. The dash-dotted line represents the \fc(\HI) at that threshold in the IGM for a velocity window of $100$~\kms. Other details are as in the {\tt left} panel.}
    \label{fig:fc_env_rho}
\end{figure*}

Fig.~\ref{fig:fc} presents the \HI\ covering fraction for different threshold column densities. The filled data points represent the average \fc(\HI) measured within the impact parameter range of $16<\rho/{\rm pkpc}<320$ and a LOS velocity window of $\pm500$~\kms. The error bars indicate the $68\%$ Wilson-score confidence intervals. The \fc(\HI) for low column density \HI\ absorbers (approximately \logN~$\approx 14.5$) is close to unity. However, as the threshold column density increases, \fc(\HI) declines.

To compare these results with random regions, we calculated the \HI\ covering fractions for the IGM at similar redshifts using the following equation from \cite{wilde2021}: 
\begin{equation}
   f_{\rm random} = 1 - {\rm exp}\left( - \left\langle \frac{dn}{dz} \right\rangle \Delta z\right) 
\label{eqn: fc_ransom}
\end{equation}
where $\langle \frac{dn}{dz} \rangle$ is the average number of \HI\ absorbers with \NHI~$> N_{\rm Th}$ per unit redshift range, calculated from the best-fitting model to the CDDF from  \cite{Kim2021}. Here, $\Delta z$ is the redshift range corresponding to a velocity window of $\pm500$~\kms\ at $z\approx3.3$. The open squares in Fig.~\ref{fig:fc} represent the covering fractions for the IGM. The \HI\ covering fraction near LAEs is consistently higher than in random IGM locations. {We obtained consistent results for the other CDDF models mentioned in literature \citep[e.g., ][]{Noterdaeme2009, Rudie_2013, Altay2011}.}

The light-blue line in the plot illustrates the fractional difference in \fc(\HI) i.e., $(f_c(\HI) - f_{\rm random})/f_{\rm random}$, with its values depicted on the opposite y-axis. This ratio gradually increases for higher threshold column densities, reflecting the higher over abundance of higher column density absorbers near the LAEs compared to random regions.

We observe no notable changes in the \HI\ covering fraction around LAEs across different redshifts for any threshold \NHI. This is expected given the small redshift range ($\Delta z/(1+z_{\rm median}) \approx0.2$) for our sample. Consequently, we do not account for any redshift evolution of \fc(\HI) near LAEs in our subsequent analysis.

\subsection{Influence of galaxy environments on \fc(\HI)} 

In previous sections, we explored the distribution of \HI\ absorbers and noted their significant incidence around LAEs compared to the IGM. In this section, we will investigate how the environment of the LAEs affects the distribution of the neutral gas around them.

To examine this, we first identify LAEs that have at least one other companion LAE within $\pm500$~\kms\ in LOS velocity space within the MUSE field of view. We refer to them as galaxy ``pairs/groups''. The  remaining LAEs are considered ``isolated''. Based on this simple classification, we find that 33 out of the total 96 LAEs are part of a pair/group.

The left panel of Fig.~\ref{fig:fc_env_rho} shows the differential \HI\ covering fraction against the impact parameter for a  threshold \logN~$=15.0$ and a velocity window of $\pm500$~\kms. The dash-dotted line represents the \fc(\HI) ($\approx 0.49$) at the same threshold column density obtained for the IGM, similar to the empty squares in Fig.~\ref{fig:fc}. The first two bins cover the range of $0-100$~pkpc and $100-200$~pkpc, respectively, and the third bin covers $200-320$~pkpc. Although there is no obvious trend between \fc\ and $\rho$, it is evident that \fc(\HI) is consistently higher ($\approx1$) for the pair/group LAEs as compared to the isolated ones ($\approx0.8$). We obtain similar trends if we use a smaller velocity window of $\pm250$~\kms\ for associating LAEs and \HI\ absorption.

The right panel of Fig.~\ref{fig:fc_env_rho} shows the differential \fc(\HI) in bins of LOS velocity for $\rho \leq 320$~pkpc (median $\rho \approx 160$~pkpc). As in the left panel, a threshold \logN $=15.0$ is used here. The dash-dotted line represents the \fc(\HI) at the same threshold column density in the IGM, but for a velocity window of $100$~\kms, consistent with the $dv$-bin size used in the plot. The pairs/groups are again seen to have higher \fc(\HI) than the isolated LAEs. Although at large LOS velocities, they become consistent within the $1\sigma$ allowed uncertainties.

Finally, we confirm that using a smaller velocity window of $\pm250$~\kms\ for defining the galaxy groups yielded a consistent trend as our fiducial value of $500$~\kms. See the tables in appendix~\ref{appendixa} for the \fc(\HI) values calculated at different impact parameter bins varying the threshold \HI\ column densities and different LOS velocities for associating \HI\ absorbers with the LAEs.

\subsection{Influence of \ew\ on \fc(\HI)}

The rest frame equivalent width of \lya\ emission (\ew) is defined as: 
\begin{equation}
{\rm EW_0} = \frac{F_{\rm ly \alpha}^{\rm line}}{f_{\rm ly \alpha}^{\rm cont} \times (1+z)}~.   
\label{eq:ew0}
\end{equation}
Here, $F_{\rm ly \alpha}^{\rm line}$ is the \lya\ emission line flux, and $f_{\rm ly \alpha}^{\rm cont}$ is the UV continuum flux density at the observed \lya\ wavelength. The continuum flux density is estimated from the extrapolation of measured continuum at rest-frame $1500$~\ang\ assuming a UV continuum slope ($\beta_{\rm UV}$) of $-2.0$ \citep[]{Bouwens2014}. In our sample, for some of the LAEs, the UV continuum is detected with $>5\sigma$ significance. In case of a non-detection, we place a lower limit on \ew\ \citep[see][for details]{Muzahid_2020}.

\begin{figure}
\includegraphics[width=\linewidth]{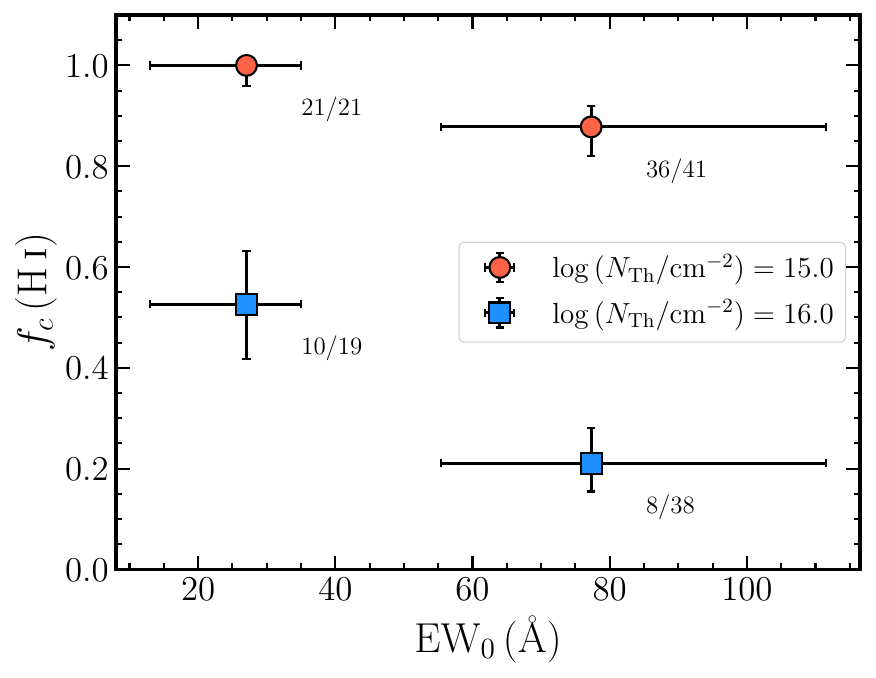}
\caption{Covering fraction as a function of the rest frame equivalent width (\ew) of the \lya\ emission. Different colors represent different threshold column densities, as indicated by the legends. The error bars along the y-axis indicate the $68\%$ Wilson-score confidence interval. The numbers adjacent to the data points correspond to the respective values of $n_{\rm Hit}$ and $ n_{\rm tot}$ (see eqn.~\ref{eq:fc}). Along the x-axis, the points represent the median values and the error bars are the $16^{\rm th}-84^{\rm th}$ percentile ranges of \ew\ in each bin. While the \fc(\HI) for the two \ew\ bins are consistent within the error bars for the lower threshold of \logN~$=15.0$, a significant anti-correlation between \fc(\HI) and \ew\ is present for the threshold \logN~$=16.0$. 
}
\label{fig:fc_ew0}
\end{figure}

Fig.~\ref{fig:fc_ew0} shows \fc(\HI) within $\pm 500$~\kms as a function of \ew\ for two different threshold \NHI\ values, as indicated in the legends. To divide the LAEs into two \ew\ bins, we first exclude 15 LAEs for which the continuum is blended with foreground sources. Next, we separate the remaining 81 non-blended LAEs based on their median \ew\ value ($51.8$~\ang). Subsequently, we exclude the lower limits on \ew\ from the lower \ew\ bin, as their values may not fall within that specific bin. However, this is not the case for LAEs with lower limits on \ew\ falling in the upper \ew\ bin, thus they are considered. Although the \fc(\HI) values for the two \ew\ bins are consistent within the error bars for the lower threshold of \logN~$=15.0$, a significant decrease of the covering factor with \ew\ is evident for the threshold of \logN~$=16.0$. We find a similar trend using a smaller velocity window of $\pm250$~\kms\ for associating absorbers with the LAEs. See appendix~\ref{appendixa} for the tabulated values of \fc(\HI). Additionally, we noted that the median value of the total \NHI\ within $\pm500$~\kms\ of the LAEs contributing to the lower \ew\ is \logN~$=16$, whereas for those contributing to the higher \ew\ bin, it is \logN~$=15.5$.  


We verified that almost $50\%$ of the LAEs in each \ew\ bin reside in a pairs/group environment. Therefore, it is unlikely that the observed trend is driven by environmental differences.



\section{Discussion}
\label{sec:discussions}

\subsection{The host galaxies of strong \HI\ absorbers}
\label{sec:discussion_LLS}

\begin{figure}
\includegraphics[width=\linewidth]{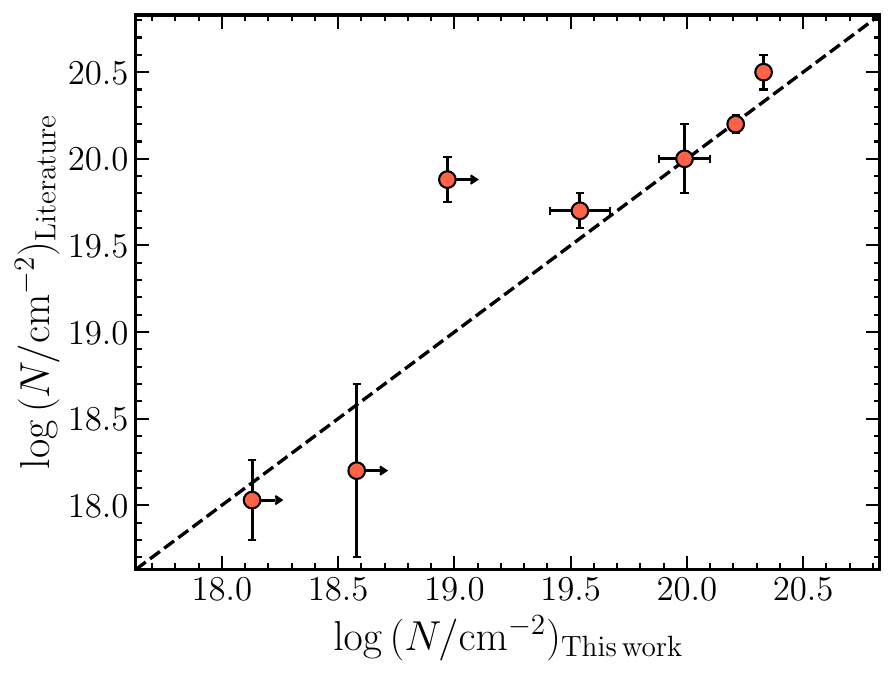}
\caption{Comparison of the \HI\ column densities measured in this work with those reported in the literature for the same systems. The dashed line indicates the 1:1 relationship. For the majority of the cases, they are in good agreement with each other.}
\label{fig:LLS_NHI_compare}
\end{figure}

The strong \HI\ absorbers are categorized as Lyman-limit systems (LLS), sub-damped Ly$\alpha$ absorbers (sub-DLAs), and DLAs, depending on their column densities (\NHI) being $>10^{17.2}$, $>10^{19.0}$, and $>10^{20.3}$ \sqcm, respectively. Among the 11 optically thick (i.e., \NHI$>10^{17.2}\, \rm cm^{-2}$) absorbers in our absorber catalog, only one is classified as a DLA and four are sub-DLAs. In Fig.~\ref{fig:LLS_NHI_compare}, we present a comparison of our estimated \NHI\ values with those from the literature for the same systems, when available \citep[]{Neeleman_2013, Prochaska_2015, Lofthouse_2023}. The plot indicates a good agreement between our measurements and the reported values in the majority of cases. Although, one absorber deviates significantly from the 1:1 line, it is actually consistent due to it being a lower limit.

High-column density \HI\ absorbers, particularly DLAs, are considered effective probes for studying gas within a few tens of kpc of galaxies \citep[e.g.,][]{Cooke2010,Rahmati_schaye2014,Rubin2015}. Pre-MUSE surveys utilizing instruments with a small FoV, such as VLT-SINFONI \citep[FoV of $10'' \times 10''$;][]{Peroux_2012} or long-slit observations using X-Shooter at multiple position angles \citep[]{Fynbo_2010, Rahmani2016, Krogager_2017}, have successfully detected DLA host galaxies, often within a few tens of kpc. Several of these earlier studies have targeted high-metallicity DLAs, as their host galaxies are likely to adhere to a mass-metallicity relation, followed by the observed metallicity-velocity relation for DLAs \citep[]{Ledoux06,Neeleman_2013}. Assuming that luminosity scales with stellar mass, it is expected that high-metallicity DLAs would have brighter galaxy counterparts, making them easier to detect.

Of the 96 LAEs in our ``absorption-blind'' sample, only one is associated with a DLA, with an impact parameter of $230$~pkpc. The LAEs associated with the four sub-DLAs have impact parameters ranging from $160$ to $240$~pkpc. These findings are in line with the MUSE-based, absorption-centric study by \citet{Mackenzie_2019}. They identified 14 LAEs associated with 6 metal-poor ($\log_{10}\, (Z/\rm Z_{\odot}) \lesssim -1$) DLAs. All but two LAEs were found to be located at $\rho > 50$~pkpc. Explaining such high neutral gas column densities at these large distances remains challenging, particularly in light of studies estimating the radius of DLAs to be $\lesssim 10$~kpc \citep[e.g.,][]{Cooke2010, Cooke2015, Rubin2015}. The large DLA-LAE separation suggests that they may not be directly linked to the absorbing gas.


All the LAEs associated with LLSs in MUSEQuBES have impact parameters of $>50$~kpc with no significant trend between \NHI\ and $\rho$ (see Fig.~\ref{fig:col_profile}). Recently, as part of the MUSE Analysis of Gas around Galaxies (MAGG) survey, \cite{Lofthouse_2023} also reported no observed trend between the \NHI\ of their LLS sample and the impact parameter of LAEs detected within $\pm1000$~\kms\ out to  $\approx 300$~pkpc. They presented a detailed discussion on the connection between optically thick gas and LAEs in their sample (see their section 7.2). They suggested that LLSs originate primarily either from the outer CGM (defined as $3-4$~$R_{\rm vir}$)  or in the IGM but in proximity to galaxies. When multiple LAEs are detected near an LLS, they observe a preferential alignment of the LAEs, which is suggestive of the fact that LAEs and optically thick absorbers may lie within filamentary structures. Such a filamentary structure is also evident in the MUSEQuBES survey for the highest LAE overdensity system detected towards Q~1317--0507 (Banerjee et al., In preparation).



Exploiting $\approx 140$h of MUSE observations, \cite{Bacon2021} found that extended diffuse \lya\ emission around LAEs at $z \approx 3.1 - 4.5$ originates from gas beyond their CGM, likely driven by a large population of ultra-low-luminosity LAEs with $L_{\rm Ly\alpha} <~10^{40}\, \rm erg\, s^{-1}$. In comparison, the median $L_{\rm Ly\alpha}$ for our sample, i.e., $ \approx 10^{42}~ \rm erg\, s^{-1}$, is two orders of magnitude higher. Those faint LAEs will remain undetected in MUSEQuBES with on-source exposure times of $\lesssim 10$~h per field, but could potentially give rise to strong \HI\ absorption in background quasar spectra. This association of strong HI absorbers with ultra-low luminosity galaxies agrees with  predictions from cosmological simulations \citep{Rahmati_schaye2014} and provides a plausible explanation for the presence of high \NHI\ absorbers at larger impact parameters in MUSE surveys.

Finally, \cite{Kusakabe2020} showed that at $z\approx 3-6$, only up to $30\%$ of star-forming galaxies are LAEs. This indicates that surveys focusing solely on LAEs may not capture the full galaxy population. Consequently, the non-detection of the true hosts of the absorbers can result in a scatter in the distribution of \NHI\ with impact parameters, as shown in Fig.~\ref{fig:col_profile}, leading to the observed lack of trend.

\color{black}

\subsection{\HI\ covering fractions} 

The characteristics of \HI\ absorption are primarily influenced by three main factors: (a) the Hubble expansion; (b) changes in the ionizing UV background radiation field (UVB) resulting from star-forming galaxies and AGN; and (c) the growth of large-scale structure. \citet{Dave_1999} demonstrated that, as a consequence of these factors, the overdensity associated with a fixed \NHI\ decreases with increasing redshift \citep[see also ][]{Theuns1998, Schaye_2001}. Consequently, absorbers with the same \NHI\ at different redshifts correspond to different types of cosmic structures. Using equation 10 from \citet{Schaye_2001}, it can be inferred that \logN~$=14.0$ corresponds to an overdensity of $\delta \approx 15$ at $z\approx 0$, whereas at $z\approx 3$, it corresponds to an overdensity of $\delta \approx 1.4$ ($(1 + \delta) =\rho/ \bar{\rho}$, where $\bar{\rho}$ is the cosmic mean matter density). These findings result in a variation in the $dn/dz$ of \HI\ absorbers at different cosmic epochs \citep[e.g.,][]{Kim2021}. This evolution of the \HI\ number density is expected to result in a higher \HI\ covering fraction at higher redshift for a fixed comoving impact parameter and even stronger effect for a fixed proper impact parameter provided collisional ionization is unimportant.



\begin{figure}
\includegraphics[width=\linewidth]{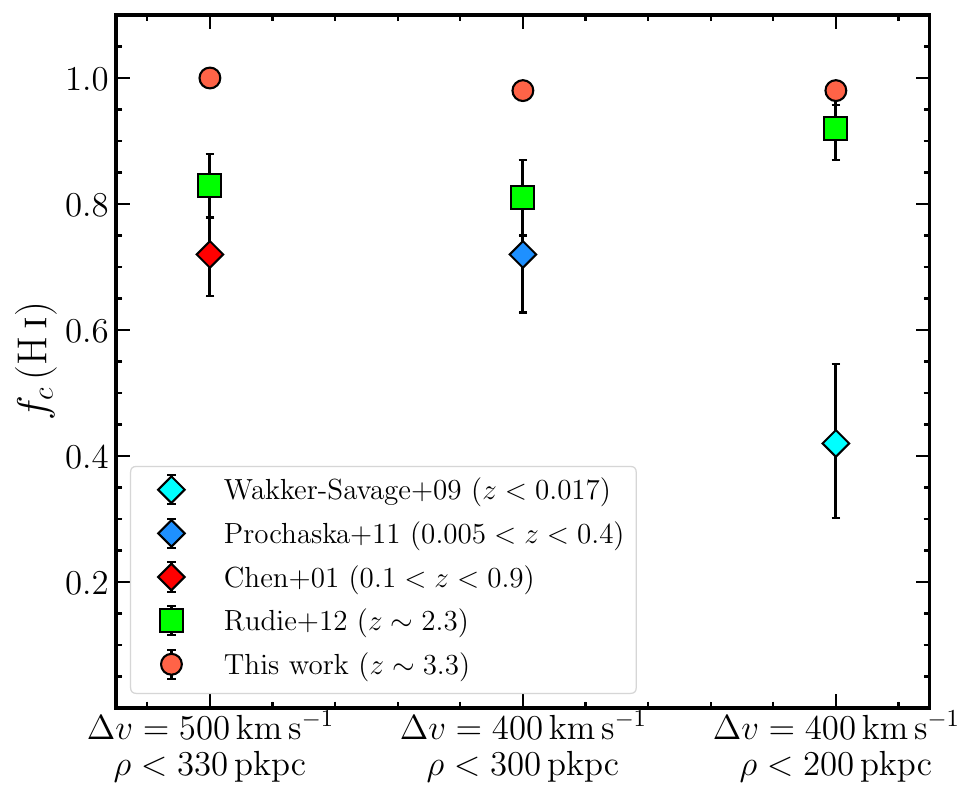}
\caption{A comparison of \HI\ covering fraction for the MUSEQuBES sample (represented by red circles) and several samples from the literature (indicated by the legends) at different redshifts, calculated at the same threshold \logN$~=14.0$. The literature samples  consist of the high-$z$ KBSS-LBGs from \citet{Rudie_2012}, low-$z$ $L_B \gtrsim 0.25L_B^*$ galaxies from \citet{Chen_2001_b}, and low-$z$ $L>0.1 L^*$ galaxies from \citet{Prochaska_2011} and \citet{Wakker-Savage_2009}. Along the x-axis, we indicate the respective LOS-velocity windows and physical impact parameter ranges used for the \fc(\HI) measurements. The error bars shown for the KBSS-LBG data are directly taken from \citet{Rudie_2012}. For the remaining cases, the error bars represent the $68\%$ Wilson-score confidence intervals. For the MUSEQuBES sample, the error bars are smaller than the circle size. MUSEQuBES consistently exhibits the highest \fc(\HI) compared to the other samples.  
}
\label{fig:fc_compare}
\end{figure}

In Fig.~\ref{fig:fc_compare} we compare the \fc(\HI) at threshold \logN$=14.0$, that we measure for MUSEQuBES with a few surveys from the literature: the KBSS-LBGs at $z\approx 2.3$ from \citet[][see their table 6]{Rudie_2012}, the $L_B \gtrsim 0.25L_B^*$ galaxies from \citet[][\fc(\HI) $= 34/47$; $0.1<z<0.9$]{Chen_2001_b}, $L>0.1 L^*$ galaxies from \citet[][\fc(\HI) $= 18/25$; $0.005<z<0.4$]{Prochaska_2011}, and \citet[][\fc(\HI) $= 6/14$; local universe]{Wakker-Savage_2009}. Along the x-axis, we indicate the LOS velocity windows and impact parameter ranges used for the respective \fc(\HI) calculations. The MUSEQuBES LAEs have the highest \fc(\HI) in all cases. Notably, it becomes comparable to the $z\approx 2.3$ LBGs for $\rho<200$~pkpc and $\Delta v = 400$~\kms. If instead of the proper impact parameter, if we calculate the \fc(\HI) of the MUSEQuBES sample for the same co-moving impact parameter as these low-$z$ surveys, it reaches $100\%$ for all three different cuts explored here.

Note that the overdensity ($\approx15$) an absorber of $N(\HI)\, \approx 10^{14.0}$~\sqcm\ at $z\approx 0$ traces, would be traced by an absorber of $N(\HI)\, \approx 10^{15.6}$~\sqcm\ at $z\approx 3.3$. The \fc(\HI) we obtained for the threshold $N(\HI)$ of $10^{15.6}$~\sqcm\ is $0.38$ within 200~pkpc and $0.62$ within 320~kpc from our survey. It is intriguing to note that these values are consistent with low-$z$ measurements of \citet{Wakker-Savage_2009} for $<200$~pkpc and \citet{Chen_2001_a,Prochaska_2011} for $\leq 330$~pkpc for threshold \logN~$14.0$, as can be seen from Fig.~\ref{fig:fc_compare}. 



\citet{Johnson2015} found \fc(\HI)$\approx 90\%$ at threshold \logN~$=14$ within the virial radii of $L>0.1L_*$ galaxies at $z<0.4$, which reduces to only $\approx 16\%$ at larger distances. The COS-Halos survey \citep{Tumlinson2013} reported \fc(\HI)$\approx 91\%$ for \logN~$>14$ within $150$~pkpc and $|\Delta v| \approx 200$~\kms\ for $\approx L_*$ galaxies at $z\approx 0.2$. In our sample, for the same LOS velocity and impact parameter cuts, the \fc(\HI)$\approx 92\%~ (38/41)$. Thus, it appears that the outskirts ($\approx 3 R_{\rm vir}$) of the LAEs in our high-$z$ sample are as \HI-rich as the inner CGM ($<0.5 R_{\rm vir}$) of $\approx L_*$ galaxies at low $z$.

The MAGG survey measured \fc(\HI) $\approx 0.2$ within $\pm500$~\kms\ and $300$~pkpc from the $z\approx 3$ LAEs in their sample for a threshold column density \logN~$=17.2$ \citep[]{Lofthouse_2023}. We obtain \fc(\HI) $= 0.16^{+0.04}_{-0.04}$ for MUSEQuBES for the same threshold. Additionally, \citet{Lofthouse_2023} reported \fc(\HI)~$=0.29 \pm 0.03$ for group galaxies and \fc(\HI)~$=0.08 \pm 0.02$ for isolated LAEs for the same threshold $N(\HI)$. In contrast, MUSEQuBES yields \fc(\HI)~$=0.10^{+0.04}_{-0.06}$ (3/29) for group galaxies and \fc(\HI)~$=0.19^{+0.04}_{-0.05}$ (9/46) for isolated LAEs. The measurements of \fc(\HI) in both surveys use similar ranges of $\Delta v$ ($\pm 500$~\kms) and $\rho$ ($\approx 300$~pkpc). Recall that only 11 absorbers associated with the LAEs in our sample have $N(\HI)>10^{17.2}$~\sqcm. We could not confirm the enhancement of \fc(\HI) for group galaxies as reported by \citet[]{Lofthouse_2023} owing to the scarcity of LLS in our study (their survey had more than 50 absorbers with \logN~$\geq17.2$). Nevertheless, we do see a strong environmental effect on the \fc(\HI) at a smaller threshold of \logN$=15.0$ out to $\approx 250$~pkpc (see Fig.~\ref{fig:fc_env_rho}).


\begin{figure}
\includegraphics[width=\linewidth]{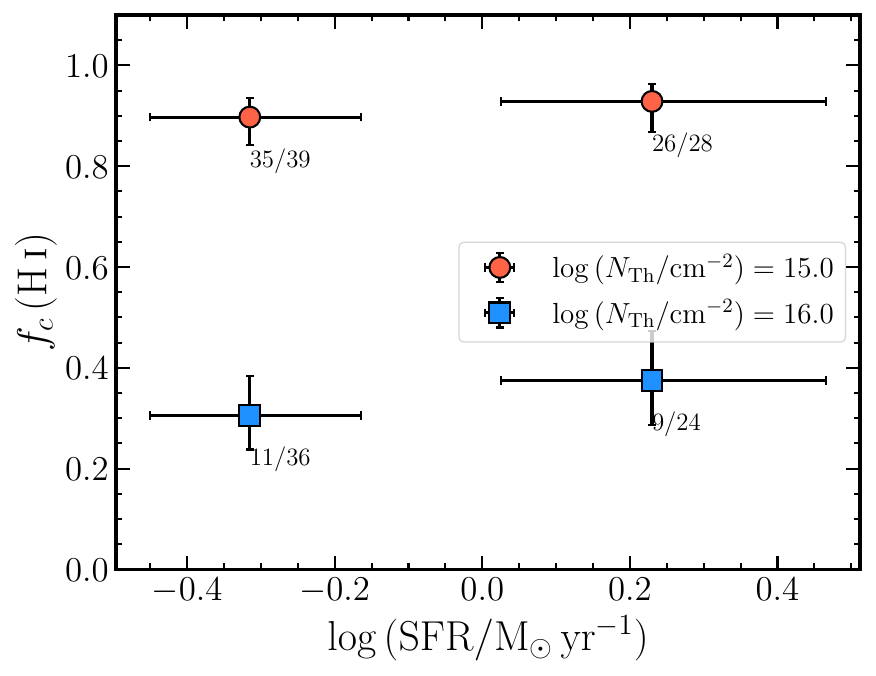}
\caption{\HI\ covering fraction (within $320$~pkpc and $\pm500$~\kms\ LOS velocity of the LAEs) calculated at two different threshold column densities as indicated by the legends for two bins of SFR. The error bars along the y-axis indicate the $68\%$ Wilson-score confidence intervals. The numbers adjacent to the blocks correspond to the respective values of $n_{\rm Hit}$ and $ n_{\rm tot}$. Along the x-axis, the points represent the median values and the error bars are the $16^{\rm th}-84^{\rm th}$ percentile ranges of SFR in each bin. No significant trend is found between \fc\ and SFR of the LAEs.}
\label{fig:fc_sfr}
\end{figure}

\cite{Muzahid_2021} reported a correlation between the rest-frame equivalent width of stacked circumgalactic \lya\ absorption and the SFR (determined via the UV continuum emission) of the MUSEQuBES LAEs. They split the LAEs into three subsamples: two subsamples with high and low SFRs; and an additional subsample encompassing LAEs for which the UV continuum was not detected with $>5\sigma$ significance, i.e., the non-detections. Their study revealed that the higher SFR subsample has a mean \HI\ absorption-equivalent width $\approx 1.6$ times larger than that of the lower SFR subsample. However, from our individual Voigt profile analysis, we did not find any significant correlation between the total \NHI\ and SFR for these LAEs. The median total \NHI\ for the higher SFR sub-sample ($10^{15.7}$\sqcm) is similar to that of the lower SFR sub-sample ($10^{15.5}$~\sqcm).

This lack of a trend is also evident in Fig.~\ref{fig:fc_sfr}, where we plot \fc(\HI) against $\rm log_{10}$~(SFR). The data are binned based on the median value of the SFR of the LAEs. We excluded the LAEs with SFR non-detections (i.e., upper limits) from the higher SFR bin but considered the LAEs when their upper limits fall in the lower SFR bin. The \fc(\HI) for LAEs in both the SFR bins, calculated for two different threshold column densities, are consistent within the error bars, indicating a lack of any significant difference in the \HI\ distribution in the CGM for higher and lower SFR subsamples. Note that, the median SFRs of these two bins only differ by a factor of $\sim 3$ with each other.



Additionally, we calculated the column density-weighted LOS velocity of \HI\ absorbers from the respective host-LAEs ($\Delta v_{\rm wgt}$), obtaining a median value of $4.8$~\kms\ and $\sigma = 175.4$~\kms. However, we do not find any correlation between the impact parameter and $\Delta v_{\rm wgt}$ as indicated by the Spearman correlation test results: $\rm r_{corr}=0.01$ and $p$-value$=0.91$. We performed this correlation test by creating smaller subsamples of galaxies based on their environments (i.e., pairs/group or isolated LAEs) and did not find any significant correlations in any of the cases.


\subsection{The relation between neutral gas in the ISM and CGM}
\label{sec:discussion_ew}

In Fig.~\ref{fig:fc_ew0}, we observed an anti-correlation between \fc(\HI) and \ew. This trend is particularly evident for the threshold \logN~$=16.0$, where the difference in \fc(\HI) between the two \ew\ bins is significant at the $\approx 2\sigma$ level. Since $\approx 50\%$ of LAEs in both the \ew\ bins are associated with the pair/group subsample, it is unlikely that environmental effects are driving the trend.

Simple shell-models of \lya\ radiative transfer predict a strong anti-correlation between $N(\HI)$ and \ew\ \citep[]{Verhamme2015, Verhamme2017}. This is because as $N(\HI)$ increases, \lya\ photons experience more scattering, unlike continuum photons. Hence, \lya photons cover a longer distance in order to shift their frequencies off-resonance to escape the ISM of the galaxies, which increases their probability of being destroyed by dust grains. Thus, the observed anti-correlation between \fc(\HI) and \ew\ could imply a positive trend between the covering fraction of neutral gas in the CGM and $N(\HI)$ in the ISM of the LAEs. In other words, galaxies whose environments are abundant in cool gas are also inherently gas-rich.  

The shell model \citep[]{Verhamme2006} also predicts that galaxies with lower \NHI\ exhibit narrower line profiles, and vice versa. However, we did not observe a significant difference in the distribution of FWHM of \lya\ emission lines between the high and low-\ew\ samples ($p_{\rm KS} = 0.2$). Both samples have a median FWHM of $\approx250$~\kms. 

An alternative explanation for the observed trend could be enhanced photoionization from the local radiation fields of LAEs with relatively higher \ew, which eventually leads to a lower \HI\ covering fraction \citep[e.g.,][]{Matthee2022}.



\subsection{Correlation between \lya\ transmission and LAEs}
\label{sec:discussion_CCF}


\begin{figure}
	\includegraphics[width=\linewidth]{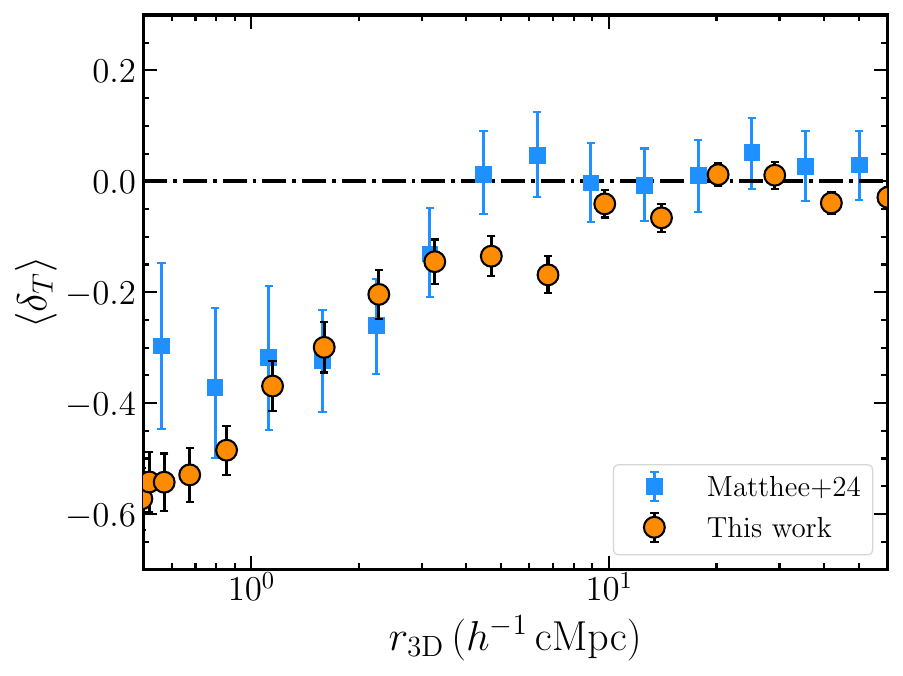}
    \caption{The mean excess-\lya-transmission ($\langle \delta_T \rangle$) as a function of 3D distance. The $\langle \delta_T \rangle$ values from this work (in orange) are compared with those at $z\approx 4.3$ from \citet{Matthee2024} (in blue). The error bars for our sample represent the $1 \sigma$ bootstrap error, calculated by resampling the LAE sample $1000$ times. The black dot-dash line indicates $\langle \delta_T \rangle=0$. 
    }
    \label{fig:T_lya}
\end{figure}

In the current section, we discuss the influence of LAEs on the \lya\ transmission as measured in the quasar spectra. We define the excess \lya\ transmission ($\delta_T$) with respect to the average transmission at a given redshift (i.e., $\langle T_z \rangle$) as:
$$\delta_T = \frac{T}{\langle T_z \rangle} - 1~.$$ 
Here, $T$ represents the normalized spectral flux.

{In Fig.~\ref{fig:T_lya}, we present the mean excess \lya\ transmission ($\langle \delta_T \rangle$: shown in orange) as a function of the 3D distance, $\Delta r_{\rm 3D}$, where $r_{\rm 3D} = \sqrt{r^2_{\parallel} + r^2_{\bot}}$. Here, $r_{\parallel}$ and $r_{\bot}$ represent the LOS and transverse distances with respect to the LAEs, respectively. To calculate this, we summed the total flux from the normalized quasar spectra within the LOS velocity window $\Delta v$ around the systemic redshift of each LAE and averaged it over the total number of pixels contributing to the respective $\Delta v$ bin considering all LAEs. We converted $\Delta v$ to $r_{\parallel}$ (in comoving Mpc or cMpc) using $\frac{\Delta v}{H(z)} \times (1+z)$, assuming a redshift of $z=3.3$, the median for these LAEs. Additionally, we assumed $r_{\bot} \approx 0.7$~cMpc, corresponding to the median impact parameter of 160~pkpc for our sample.} 
To model the redshift evolution of $\langle T_z \rangle$ in our data, we followed a method similar to \citet[][see also \citet{Becker_2015}]{Matthee2024}. We used the following polynomial function that best fits our data:
$$\langle T_z \rangle = -1.30 + 1.34\, z - 0.23\, z^2~.$$ 

We have also shown the $\langle \delta_T \rangle$ near the LAEs at a slightly higher redshift of $\approx 4.3$, as found by \citet{Matthee2024} (in blue). The data points are mostly within each other's error bars.

\section{Summary and conclusions}
\label{sec:summary}

This work is a part of the MUSEQuBES survey at high-$z$ \citep[]{Muzahid_2020, Muzahid_2021}, which focuses on the CGM of high $z$ LAEs. The MUSEQuBES sample consists of 96 LAEs (median \lya\ luminosity $\approx 10^{42}\, \rm erg\, s^{-1}$ and a median SFR of $\approx 1.3$~\Msun~$\rm yr^{-1}$, without accounting for dust extinction) within the redshift range $2.9$--$3.8$ that were detected in $8$ MUSE fields centered on $8$ bright background quasars. The high-resolution optical spectra of these $8$ quasars obtained with VLT/UVES and/or Keck/HIRES enabled us to explore the neutral hydrogen (\HI) absorption arising near these 96 galaxies.

We estimated the systemic redshifts of the LAEs using the empirical relation given in \citet{Muzahid_2020}. Subsequently, we conducted a Voigt profile decomposition of the Lyman series lines detected within a velocity range of $\pm500$~\kms\ around the LAEs. The column densities of the 800 \HI\ absorption components observed in these spectra spans a range from \logN\ $=11.7 - 20.3$, with a median value of \logN\ $=13.7$. In cases where we were unable to precisely constrain the column density due to blending or the absence of higher-order unsaturated Lyman series lines, we reported the lower limit on the column density. Our main findings are:\\

\begin{itemize}
    
    \item We do not find any significant correlation between the total \NHI\ and impact parameter within 16--315~pkpc (median $165$~pkpc; Fig.~\ref{fig:col_profile}). \Rev{However, the total \NHI\ within $\pm500$~\kms\ of the LAEs is generally higher than the typical \NHI\ expected from the IGM at similar redshifts. This trend also holds for absorbers within $\pm250$~\kms\ of the LAE redshifts, suggesting that \HI\ absorption is elevated near the LAEs compared to the IGM.} 
    
    \item {All the LAEs associated with \HI\ absorbers of \logN~$>17.2$ are detected at impact parameters $>50$~pkpc (Fig.~\ref{fig:col_profile}). This is likely due to the non-detection of the actual absorber hosts owing to their faintness, as discussed in Section \ref{sec:discussion_LLS}}.
    
    \item  The covering fraction, \fc(\HI), is consistently higher near the LAEs as compared to random regions for thresholds \logN~$=14-16$,  with $\rho \leq 320$~pkpc. The relative enhancement in \fc(\HI) with respect to random regions increases with the threshold column density (Fig.~\ref{fig:fc}).  

    \item The pairs/group galaxies exhibit a higher \fc(\HI) compared to isolated LAEs for a threshold \logN$=15$ along transverse directions (Fig.~\ref{fig:fc_env_rho}: {\it left}). The pairs/group LAEs display a $100\%$ \HI\ covering fraction, extending up to $\approx 250$~pkpc. 

    \item  The \fc(\HI) decreases with LOS velocity between the LAEs and absorbers (Fig.~\ref{fig:fc_env_rho}, {\it right}). The trend of pairs/group having higher \fc(\HI) compared to isolated ones is also visible along the LOS-direction, although only upto $\approx \pm300$~\kms, beyond which \fc(\HI) values are comparable within $1\sigma$.  
    
    \item For a threshold \logN~$=16$, \fc(\HI) anti-correlates with the rest frame equivalent width (\ew) of \lya\ emission (Fig.~\ref{fig:fc_ew0}). According to the \lya\ shell model, a lower \ew\ results from a large reservoir of neutral gas present within the galaxies. This could imply that galaxies residing in cool-gas-rich environments are also inherently gas-rich. The other reason might be that high-\ew\ LAEs are efficient in ionizing their surrounding medium which decreases the \fc(\HI) (see Section~\ref{sec:discussion_ew}). 

    \item Finally, we compared the excess \lya\ transmission, $\langle \delta_T \rangle$, from our observation and that of the LAEs at slightly higher redshift of $4.3$ and found they are mostly consistent with each other.

\end{itemize}

This study presents the connection between $z\approx 3$ LAEs and cool neutral gas in their surroundings, for the first time, in an ``absorption-blind'' manner using detailed Voigt profile decomposition of \HI\ absorbers. In the future, we will present ionization models to derive the density and metallicity of these absorbers and investigate their relation with the LAEs.




\section*{ACKNOWLEDGEMENT}

We would like to thank the anonymous referee for useful comments. We thank Marijke Segers, Lorrie Straka, and Monica Turner for their early contributions to the MUSEQuBES project. We thank Raghunathan Srianand for useful suggestions. EB thanks Labanya Kumar Guha and Yucheng Guo for helpful discussions. We gratefully acknowledge the European Research Council (ERC) for funding this project through the Indo-Italian grant. This paper uses the following software: NumPy \cite[]{Harris_2020}, SciPy \cite[]{Virtanen_2020}, Matplotlib \cite[]{hunter_2007}, and AstroPy \cite[]{Astropy_2013, Astropy_2018}.


\bibliographystyle{mnras}
\bibliography{zbib} 



\appendix
\label{appendixa}

\begin{table}[]
\centering
\caption{\fc(\HI) at threshold $\log N(\rm H\, {\textsc i})/ cm^{-2}=15$ for different Impact parameters ($\rho$):}
\begin{tabular}{| *{5}{c} |}
\hline
Impact parameter  & \multicolumn{2}{c|}{$\Delta v= 500$~\kms} & \multicolumn{2}{c|}{$\Delta v= 250$~\kms}\\
\hline
 (pkpc)  &   Pairs/Group  &   Isolated  &   Pairs/Group &   Isolated   \\
\hline
0 -- 100   &    $1.0_{-0.18}\, (4/4)$  &   $0.81^{+0.08}_{-0.13}\, (9/11)$  &    $1.0_{-0.18}\, (4/4)$  &   $0.67^{+0.11}_{-0.13}\, (8/12)$  \\
   
100 -- 200   &   $1.0_{-0.05}\, (17/17)$  &   $0.83^{+0.06}_{-0.08}\, (24/29)$  &    $0.82^{0.07}_{-0.1}\, (14/17)$  &   $0.73^{+0.07}_{-0.08}\, (22/30)$    \\
   
200 -- 320   &  $1.0_{-0.07}\, (12/12)$     &   $0.79^{+0.07}_{-0.10}\, (15/19)$    &    $0.83^{0.08}_{-0.12}\, ()10/12$  &   $0.44^{+0.11}_{-0.10}\, (8/18)$ \\

$\leq 320$   &  $1.0_{-0.02}\, (33/33)$     &   $-0.81^{+0.04}_{-0.05}\, (48/59)$   &    $0.84^{0.05}_{-0.07}\, (28/33)$  &   $0.63^{+0.06}_{-0.06}\, (38/60)$  \\
\hline
\end{tabular}
\\ Note: We have included $\frac{n_{\rm Hit} (N \geq N_{\rm Th})}{n_{\rm Total}}$ values in parentheses beside each entry.
\label{appen_tab:fc_rho_15}
\end{table}

\begin{table}[]
\centering
\caption{\fc(\HI) at threshold $\log N(\rm H\, {\textsc i})/ cm^{-2}=16$ for different Impact parameters ($\rho$):}
\begin{tabular}{| *{5}{c}| }
\hline
Impact parameter  & \multicolumn{2}{c|}{$\Delta v= 500$~\kms} & \multicolumn{2}{c|}{$\Delta v= 250$~\kms}\\
\hline
(pkpc)  &   Pairs/Group  &   Isolated  &   Pairs/Group &   Isolated   \\
\hline
0 -- 100   &    $0.5^{0.21}_{-0.21}\, (2/4)$  &   $0.33^{+0.16}_{-0.12}\, (3/9)$  &    $0.5^{+0.21}_{-0.21}\, (2/4)$  &   $0.3^{+0.15}_{-0.11}\, (3/10)$  \\
   
100 -- 200   &   $0.36^{0.13}_{-0.10}\, (5/14)$  &   $0.39^{+0.09}_{-0.08}\, (11/28)$  &    $0.28^{0.12}_{-0.1}\, (4/14)$  &   $0.37^{+0.10}_{-0.08}\, (11/29)$    \\
   
200 -- 320   &  $0.45^{0.14}_{-0.13}\, (5/11)$     &   $0.35^{+0.11}_{-0.10}\, (6/17)$    &    $0.36^{0.14}_{-0.12}\, (4/11)$  &   $0.22^{+0.10}_{-0.08}\, (4/18)$ \\

$\leq 320$   &  $0.41^{0.09}_{-0.08}\, (12/29)$     &   $0.37^{+0.06}_{-0.06}\, (20/54)$   &    $0.34^{0.09}_{-0.07}\, (10/29)$  &   $0.31^{+0.06}_{-0.05}\, (18/57)$  \\
\hline
\end{tabular}
\\ Note: We have included $\frac{n_{\rm Hit} (N \geq N_{\rm Th})}{n_{\rm Total}}$ values in parentheses beside each entry.
\label{appen_tab:fc_rho_16}
\end{table}

\begin{table}[]
\centering
\caption{\fc(\HI) at for different LOS velocity window ($\rho \leq 320$~pkpc):}
\begin{tabular}{| *{5}{c} |}
\hline
LOS velocity & \multicolumn{2}{c|}{Threshold $\log N(\rm H\, {\textsc i})/ cm^{-2}=15$} & \multicolumn{2}{c|}{Threshold $\log N(\rm H\, {\textsc i})/ cm^{-2}=16$}\\
\hline
 (\kms)  &   Pairs/Group  &   Isolated  &   Pairs/Group &   Isolated   \\
\hline
0 -- 100   &    $0.61^{+0.08}_{-0.08}\, (20/33)$  &   $0.39^{+0.06}_{-0.06}\, (23/59)$  &  $0.26^{+0.08}_{-0.07}\, (8/30)$  &   $0.12^{+0.05}_{-0.03}\, (7/57)$ \\
   
100 -- 200   &   $0.45^{+0.08}_{-0.08}\, (15/33)$  &   $0.28^{+0.06}_{-0.05}\, (17/60)$  &   $0.06^{+0.05}_{-0.03}\, (2/32)$  &   $0.17^{+0.05}_{-0.04}\, (10/59)$ \\
   
200 -- 300   &  $0.24^{+0.08}_{-0.06}\, (8/33)$     &   $0.12^{+0.05}_{-0.03}\, (7/57)$    &  $0.0^{+0.03}_{-0.0}\, (0/31)$  &   $0.07^{+0.04}_{-0.02}\, (4/57)$ \\

300 -- 400   &  $0.24^{+0.08}_{-0.06}\, (8/33)$     &   $0.16^{+0.05}_{-0.04}\, (9/56)$    &  $0.15^{+0.07}_{-0.05}\, (5/33)$  &   $0.02^{+0.03}_{-0.01}\, (1/55)$ \\

400 -- 500   &  $0.18^{+0.07}_{-0.06}\, (6/32)$     &   $0.14^{+0.05}_{-0.04}\, (9/61)$    &  $0.06^{+0.05}_{-0.03}\, (2/31)$  &   $0.03^{+0.03}_{-0.02}\, (2/60)$ \\
\hline
\end{tabular}
\\ Note: We have included $\frac{n_{\rm Hit} (N \geq N_{\rm Th})}{n_{\rm Total}}$ values in parentheses beside each entry.
\label{appen_tab:fc_dv}
\end{table}

\begin{table}[]
\centering
\caption{\fc(\HI) for different \ew:}
\begin{tabular}{| *{5}{c} |}
\hline
\ew\  & \multicolumn{2}{c|}{Threshold $\log N(\rm H\, {\textsc i})/ cm^{-2}=15$} & \multicolumn{2}{c|}{Threshold $\log N(\rm H\, {\textsc i})/ cm^{-2}=16$}\\
\hline
 (\ang)  &   $\Delta v= 500$~\kms  &   $\Delta v= 250$~\kms  &   $\Delta v= 500$~\kms &   $\Delta v= 250$~\kms  \\
\hline
$< 52$   &    $1.0_{-0.04}\, (21/21)$  &   $0.86^{+0.05}_{-0.08}\, (19/22)$  &   $0.52^{+0.10}_{-0.10}\, (10/19)$  &   $0.4^{+0.10}_{-0.10}\, (8/20)$  \\
   
$> 52$   &  $0.87^{+0.04}_{-0.06}\, (36/41)$  &   $0.70^{+0.06}_{-0.07}\, (29/41)$  &   $0.21^{+0.07}_{-0.05}\, (8/38)$  &   $0.21^{+0.07}_{-0.05}\, (8/38)$   \\
   
\hline
\end{tabular}
\label{appen_tab:fc_ew}
\\ Note: 52~\ang\ is the median \ew\ for our sample. Also, we have included $\frac{n_{\rm Hit} (N \geq N_{\rm Th})}{n_{\rm Total}}$ values in parentheses beside each entry.
\end{table}


\end{document}